\documentclass[twoside]{article}
\usepackage{epsfig}
\usepackage{rotating}
\catcode`\@=11
\long\def\@makefntext#1{
\protect\noindent \hbox to 3.2pt {\hskip-.9pt  
$^{{\eightrm\@thefnmark}}$\hfil}#1\hfill}		

\def\@makefnmark{\hbox to 0pt{$^{\@thefnmark}$\hss}}	
	
\def\ps@myheadings{\let\@mkboth\@gobbletwo
\def\@oddhead{\hbox{}
\rightmark\hfil\eightrm\thepage}   
\def\@oddfoot{}\def\@evenhead{\eightrm\thepage\hfil
\leftmark\hbox{}}\def\@evenfoot{}
\def\sectionmark##1{}\def\subsectionmark##1{}}

\oddsidemargin=\evensidemargin
\newcounter{sectionc}\newcounter{subsectionc}\newcounter{subsubsectionc}
\renewcommand{\section}[1] {\vspace{12pt}\addtocounter{sectionc}{1} 
\setcounter{subsectionc}{0}\setcounter{subsubsectionc}{0}\noindent 
	{\tenbf\thesectionc. #1}\par\vspace{5pt}}
\renewcommand{\subsection}[1] {\vspace{12pt}\addtocounter{subsectionc}{1} 
	\setcounter{subsubsectionc}{0}\noindent 
	{\bf\thesectionc.\thesubsectionc. {\kern1pt \bfit #1}}\par\vspace{5pt}}
\renewcommand{\subsubsection}[1] {\vspace{12pt}\addtocounter{subsubsectionc}{1}
	\noindent{\tenrm\thesectionc.\thesubsectionc.\thesubsubsectionc.
	{\kern1pt \tenit #1}}\par\vspace{5pt}}
\newcommand{\nonumsection}[1] {\vspace{12pt}\noindent{\tenbf #1}
	\par\vspace{5pt}}

\newcounter{appendixc}
\newcounter{subappendixc}[appendixc]
\newcounter{subsubappendixc}[subappendixc]
\renewcommand{\thesubappendixc}{\Alph{appendixc}.\arabic{subappendixc}}
\renewcommand{\thesubsubappendixc}
	{\Alph{appendixc}.\arabic{subappendixc}.\arabic{subsubappendixc}}

\renewcommand{\appendix}[1] {\vspace{12pt}
        \refstepcounter{appendixc}
        \setcounter{figure}{0}
        \setcounter{table}{0}
        \setcounter{lemma}{0}
        \setcounter{theorem}{0}
        \setcounter{corollary}{0}
        \setcounter{definition}{0}
        \setcounter{equation}{0}
        \renewcommand{\thefigure}{\Alph{appendixc}.\arabic{figure}}
        \renewcommand{\thetable}{\Alph{appendixc}.\arabic{table}}
        \renewcommand{\theappendixc}{\Alph{appendixc}}
        \renewcommand{\thelemma}{\Alph{appendixc}.\arabic{lemma}}
        \renewcommand{\thetheorem}{\Alph{appendixc}.\arabic{theorem}}
        \renewcommand{\thedefinition}{\Alph{appendixc}.\arabic{definition}}
        \renewcommand{\thecorollary}{\Alph{appendixc}.\arabic{corollary}}
        \renewcommand{\theequation}{\Alph{appendixc}.\arabic{equation}}
        \noindent{\tenbf Appendix \theappendixc #1}\par\vspace{5pt}}
\newcommand{\subappendix}[1] {\vspace{12pt}
        \refstepcounter{subappendixc}
        \noindent{\bf Appendix \thesubappendixc. {\kern1pt \bfit #1}}
	\par\vspace{5pt}}
\newcommand{\subsubappendix}[1] {\vspace{12pt}
        \refstepcounter{subsubappendixc}
        \noindent{\rm Appendix \thesubsubappendixc. {\kern1pt \tenit #1}}
	\par\vspace{5pt}}

\newcommand{\textlineskip}{\baselineskip=13pt}
\newcommand{\smalllineskip}{\baselineskip=10pt}

\def\eightcirc{
\begin{picture}(0,0)
\put(4.4,1.8){\circle{6.5}}
\end{picture}}
\def\eightcopyright{\eightcirc\kern2.7pt\hbox{\eightrm c}} 

\newcommand{\copyrightheading}[1]
	{\vspace*{-2.5cm}\smalllineskip{\flushleft
	{\footnotesize International Journal of Modern Physics A, #1}\\
	{\footnotesize $\eightcopyright$\, World Scientific Publishing
	 Company}\\
	 }}

\def\abstracts#1#2#3{{
	\centering{\begin{minipage}{4.5in}\baselineskip=10pt\footnotesize
	\parindent=0pt #1\par 
	\parindent=15pt #2\par
	\parindent=15pt #3
	\end{minipage}}\par}} 

\newcommand{\bibit}{\nineit}

\renewenvironment{thebibliography}[1]
	{\frenchspacing
	 \ninerm\baselineskip=11pt
	 \begin{list}{\arabic{enumi}.}
	{\usecounter{enumi}\setlength{\parsep}{0pt}
	 \setlength{\leftmargin 12.7pt}{\rightmargin 0pt} 
	 \setlength{\itemsep}{0pt} \settowidth
	{\labelwidth}{#1.}\sloppy}}{\end{list}}

\newcounter{itemlistc}
\newcounter{romanlistc}
\newcounter{alphlistc}
\newcounter{arabiclistc}

\newcommand{\fcaption}[1]{
        \refstepcounter{figure}
        \setbox\@tempboxa = \hbox{\footnotesize Fig.~\thefigure. #1}
        \ifdim \wd\@tempboxa > 5in
           {\begin{center}
        \parbox{5in}{\footnotesize\smalllineskip Fig.~\thefigure. #1}
            \end{center}}
        \else
             {\begin{center}
             {\footnotesize Fig.~\thefigure. #1}
              \end{center}}
        \fi}

\newcommand{\tcaption}[1]{
        \refstepcounter{table}
        \setbox\@tempboxa = \hbox{\footnotesize Table~\thetable. #1}
        \ifdim \wd\@tempboxa > 5in
           {\begin{center}
        \parbox{5in}{\footnotesize\smalllineskip Table~\thetable. #1}
            \end{center}}
        \else
             {\begin{center}
             {\footnotesize Table~\thetable. #1}
              \end{center}}
        \fi}

\def\@citex[#1]#2{\if@filesw\immediate\write\@auxout
	{\string\citation{#2}}\fi
\def\@citea{}\@cite{\@for\@citeb:=#2\do
	{\@citea\def\@citea{,}\@ifundefined
	{b@\@citeb}{{\bf ?}\@warning
	{Citation `\@citeb' on page \thepage \space undefined}}
	{\csname b@\@citeb\endcsname}}}{#1}}

\newif\if@cghi
\def\cite{\@cghitrue\@ifnextchar [{\@tempswatrue
	\@citex}{\@tempswafalse\@citex[]}}
\def\citelow{\@cghifalse\@ifnextchar [{\@tempswatrue
	\@citex}{\@tempswafalse\@citex[]}}
\def\@cite#1#2{{$\null^{#1}$\if@tempswa\typeout
	{IJCGA warning: optional citation argument 
	ignored: `#2'} \fi}}

\def\pmb#1{\setbox0=\hbox{#1}
	\kern-.025em\copy0\kern-\wd0
	\kern.05em\copy0\kern-\wd0
	\kern-.025em\raise.0433em\box0}

\def\fnt#1#2{\footnotetext{\kern-.3em
	{$^{\mbox{\scriptsize #1}}$}{#2}}}

\def\fpage#1{\begingroup
\voffset=.3in
\thispagestyle{empty}\begin{table}[b]\centerline{\footnotesize #1}
	\end{table}\endgroup}

\def\runninghead#1#2{\pagestyle{myheadings}
\markboth{{\protect\footnotesize\it{\quad #1}}\hfill}
{\hfill{\protect\footnotesize\it{#2\quad}}}}
\font\tenrm=cmr10
\font\tenit=cmti10 
\font\tenbf=cmbx10
\font\bfit=cmbxti10 at 10pt
\font\ninerm=cmr9
\font\nineit=cmti9

\font\eightrm=cmr8

\def\qed{\hbox{${\vcenter{\vbox{			
   \hrule height 0.4pt\hbox{\vrule width 0.4pt height 6pt
   \kern5pt\vrule width 0.4pt}\hrule height 0.4pt}}}$}}

\begin{document}

\runninghead{Electro-optical measurements of ultrashort 45 MeV electron beam bunch} 
{Electro-optical measurements of ultrashort 45 MeV electron beam bunch}

\normalsize\textlineskip
\thispagestyle{empty}
\setcounter{page}{1}

\copyrightheading{}			

\vspace*{0.88truein}

\fpage{1}
\centerline{\bf Electro-optical measurements of ultrashort}
\vspace*{0.035truein}
\centerline{\bf 45 MeV electron beam bunch}
\vspace*{0.37truein}
\centerline{\footnotesize D. Nikas\footnote{{\it E-mail address:}Nikas@bnl.gov 
Tel: (631) 344-4717; Fax: (631) 344-5568 }   ,V. Castillo, L. Kowalski$^a$, 
R. Larsen, D. M. Lazarus,}
\centerline{\footnotesize C. Ozben, Y. K. Semertzidis, T. Tsang, and 
T. Srinivasan-Rao}
\vspace*{0.015truein}
\centerline{\footnotesize\it Brookhaven National Laboratory, Upton, NY 11973, USA}
\centerline{\footnotesize\it $^a${Montclair State University, Upper Montclair, NJ 07043, USA}}
\baselineskip=10pt

\vspace*{0.21truein}
\abstracts{}{We have made an observation of 45 MeV electron
 beam bunches using the nondestructive electro-optical (EO) technique. 
The amplitude of the EO modulation was found to increase linearly with 
electron beam charge and decrease inversely with the optical beam path 
 distance from the electron beam. The risetime of the 
signal was bandwidth limited by our detection system to ${\rm\sim70\ ps}$. An 
EO signal due to ionization caused by the electrons traversing the EO crystal 
was also observed. The 
EO technique may be ideal for the measurement of bunch structure with 
femtosecond resolution of relativistic charged particle beam bunches.}{}


\vspace*{1pt}\textlineskip	
\section{Introduction}	
\vspace*{-0.5pt}
\noindent
Since the first EO observation$^1$ of charge particle beam we have constructed
 an optical probe based on the electro-optical Pockels effect. That is, 
when an electric field is applied to a birefringent crystal an optical 
phase shift is introduced between orthogonal components. 
To probe it, a laser beam polarized at ${\rm 45^o}$ to the z-axis 
of the EO crystal is propagated along the y-axis of the crystal. This phase 
retardation is converted to an intensity modulation by a $\lambda \over {\rm4}$ 
plate  followed by an analyzer. The intensity of light 
$I(t)$ exiting the analyzer can be described by$^2$ 
\begin{equation}
I(t) = I_{o} [ \eta\ +\ {\rm sin}^2 ( \Gamma_o\ +\ \Gamma_b\ +\ \Gamma (t))],
\label{I(t)}
\end{equation}
where $I_o$ is the input light intensity, $\eta$ the imperfection of crystal, polarizer 
and other optics, $\Gamma_o$ is the crystal residual birefringence, 
$\Gamma_b$ is the optical bias of the system which is set at $\pi \over 4$, 
and $\Gamma (t)$ is the phase induced by  the electric field on the 
crystal. For a weak modulation, $\Gamma (t) \ll 1$, the EO component can be 
written as
\begin{equation} 
[{{I (t)}\over I_o}]_{\rm EO}\ \sim\ \Gamma (t)\ =\ {1 \over 2}({n_e}^3 r_{33} - {n_o}^3 r_{13}){{2\pi L E_z (t)} \over 
\lambda}
\label{EO}
\end{equation}
\par
The optical phase shift $\Gamma (t)$ is linearly proportional to the time-dependent 
field $E_z (t)$ induced by the passage of the electron beam,
with $L=\Delta t \times {c \over n} \simeq {r \over \gamma n}$ the distance light 
travels inside the crystal in the presence of $E_z 
(t)$, $n_e$ and $n_o$ the extraordinary and 
ordinary indices of refraction and $r_{33}$, $r_{13}$ the EO coefficients.
\par
A relativistic beam produces an anisotropically directed radial field nearly
orthogonal to the beam direction and along the z-axis of the EO crystal 
with strength$^3$
\begin{equation} 
E_z (t) = {1 \over {4 \pi \epsilon_o}}{{\gamma\ N_e\ q\ T(t)} \over {\epsilon\ r^2}}
\label{Ez}
\end{equation}
where $\gamma$ is the Lorentz factor, $N_e$ the number of electrons in the 
beam, $q$ the electron charge, $T(t)$ the temporal charge distribution, $\epsilon_o$ 
the permittivity of free space, $\epsilon$ the dielectric constant of the EO crystal 
in the z-axis direction, and $r$ the radial distance of the electron beam from the 
axis of the optical beam. Finally
\begin{equation} 
[{{I (t)}\over I_o}]_{\rm EO}\ \simeq\ ({n_e}^3 r_{33} - {n_o}^3 r_{13}) 
{{N_e\ q\ T(t)} \over {4\ \lambda\ n\ \epsilon_o\ \epsilon\ r}}
\label{final-EO}
\end{equation}
\section{Experiment}
\vspace*{-0.5pt}
\noindent
A vacuum compatible EO modulator setup was constructed using discrete optical 
components. A Nd:YAG laser, emitting 250 mW of CW power at ${\rm 1.3\ \mu m}$ was coupled 
to a vacuum sealed polarization maintaining fiber collimator and the output 
was  rotated ${\rm +45^o}$ to the azimuthal. The collimated 
${\rm 0.4\ mm}$ diameter light beam, with polarization purity 
of ${\rm \sim10^-2}$, was directed to the 
${\rm LiNbO_3}$ crystal mounted on a ceramic holder that has 
a clearance hole of 6.35 mm for the electron beam. The size of the 
crystal was 6.5(L) x 2.2(H) x 1(W) mm; the optical z-axis was aligned
 azimuthally and the x-axis was parallel to the propagation 
direction of the e$^-$ beam. Fluorescent material was placed on the ceramic 
for guiding the e$^-$ beam through the EO 
crystal. A CCD camera and a ${\rm 45^o}$ pop-up flag were also 
used for electron beam measurements. The electron beam contained up to 0.6 nC 
charge with beam diameter of ${\rm\sim0.5\ mm}$ in 10 ps bunch length at a 
repetition rate of 1.5 Hz.
\par
A vacuum sealed multimode fiber collimator collected the light output 
from the analyzer and was coupled separately to 1, 12 GHz photodiode which 
were connected to digitizing oscilloscopes with bandwidth 1, 7GHz.
\par
\section{Results}
\vspace*{-0.5pt}
\noindent
The electron beam induced EO signal origin was confirmed: 
(1) The signal vanished in the absence of electron or laser beam 
(2) The signal polarity changed sign when the direction of the electrical 
field was reversed (by placing the e$^-$ beam above and bellow the crystal), or when the input laser polarization was rotated by ${\rm 90^o}$ ,
see inset of Fig.1(Left) (a),(b) respectivelly. 
Fig.1(Left) shows the measured pulse with risetime of ${\rm\sim70\ ps}$ and in dashed line is 
the instrument response to a ${\rm\sim15\ ps}$ laser pulse which shows that our measurement was bandwidth limited by the electronics.
\par
The EO signal dependence on electron beam charge was investigated.  The 
charge was measured using a Faraday cup and a stripline. 
The e$^-$ beam was clearly passing below the EO crystal unobstructed. 
A linear $\chi^2$ minimization fit to 
the signal amplitude for 5 charge values is shown in the inset of Fig.1(Right).
\begin{figure}[htbp]
\begin{center}
{\epsfig{file=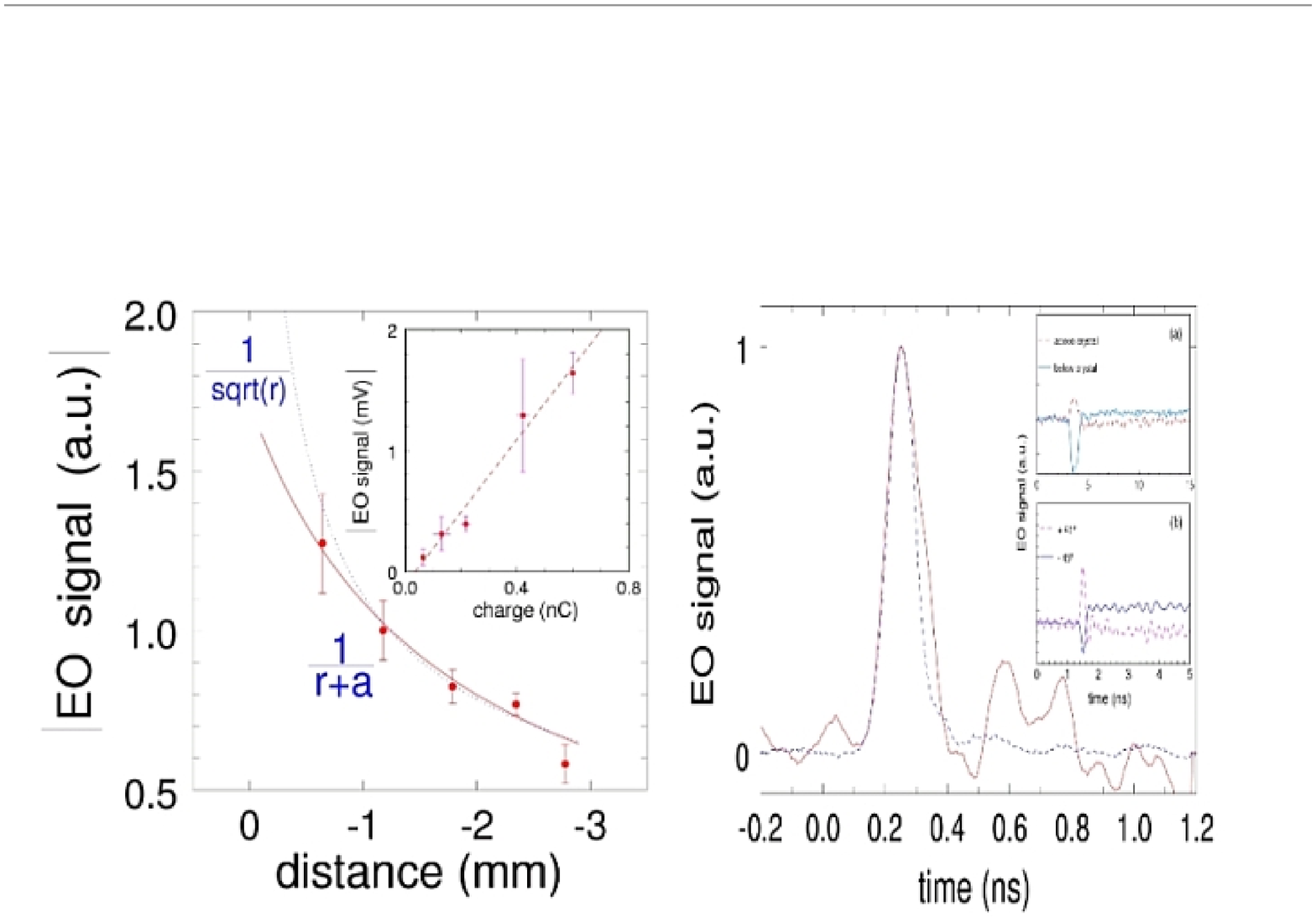, height=13cm, width=6cm, angle=-90}}
\end{center}
\caption{Left:EO signal(solid), instrument response(dashed) and polarity change(inset (a)and(b));   Right:
EO signal amplitude vs distance and charge(inset).}
\end{figure}
\par
EO signal dependence on electron beam position was also investigated. Fig.1(Right) 
displays 5 signal amplitudes when the beam was steered vertically toward, 
but not traversing the crystal, versus their distance 
from the center of the laser beam path. A $\chi^2$ minimization fit of the data 
favors a $ 1 \over {r+a}$ dependence, where $a=1.75 mm$.
\par
As the electron beam approached the optical beam path a 
distinctive positive signal with a long ${\rm\sim100\ ns}$ decay time superimposed 
on the negative EO modulation was observed which becomes negative when tranverses 
the optical beam path. It is the electron beam that ionizes the 
${\rm LiNbO_3}$ crystal creating electron-hole pairs. Since the mobility of ions is small 
compared to the electrons, an ion field remains and produces an EO signal
 opposite to that due to the electron beam field. Its decay time will be dictated 
by the electron-hole recombination time of the crystal$^4$.
\par 
\vspace*{-4pt}\textlineskip
\section{Conclusions}
\vspace*{-1.5pt}
\noindent
The effectiveness of a Pockels cell field sensor has been demonstrated for 
nondestructive measurement of an ultrashort beam bunch.
 Using an upgraded pump-probe EO detection scheme and 
state-of-the-art ultrafast optical pulse measurement techniques such as 
frequency-resolved optical gating or spectral phase interferometry for direct 
electric-field reconstruction, femtosecond electron bunch may be 
studied. Furthermore, one can in principle construct a 2-dimensional EO detector array
 to measure the spatial and temporal profile of the charged particle beam bunch.
\par
\vspace*{-5pt}
\nonumsection{References}

\end{document}